\documentclass[final,3p,twocolumn]{elsarticle}
\usepackage{amssymb}
\usepackage{amsmath}
\usepackage{graphicx}% Include figure files
\journal{Physica E}
\usepackage{tikz}

\begin{document}

\begin{frontmatter}



\title{Regular and singular Fermi liquid in triple quantum dots: Coherent transport studies}

\author[1,2]{S. B. Tooski}
\author[2,3]{A. Ram{\v{s}}ak}
\author[1]{B. R. Bu{\l}ka}\ead{bulka@ifmpan.poznan.pl}
\address[1]{Institute of Molecular Physics, Polish Academy of Sciences, ul. M. Smoluchowskiego 17, 60-179 Pozna{\'n}, Poland}
\address[2]{Jo\v zef Stefan Institute, Ljubljana, Slovenia}
\address[3]{Faculty of Mathematics and Physics, University of Ljubljana, Ljubljana, Slovenia}

\begin{abstract}
A system of three coupled quantum dots in a triangular geometry (TQD) with electron-electron interaction and symmetrically coupled to two leads is  analyzed with respect to the electron transport by means of the numerical renormalization group. Varying gate potentials this system exhibits extremely rich range of regimes with different many-electron states with various local spin orderings. It is demonstrated how the Luttinger phase changes in a controlled manner which then via the Friedel sum rule formula exactly reproduces the conductance through the TQD system. The analysis of the uncoupled TQD molecule from the leads gives a reliable qualitative understanding of various relevant regimes and gives an insight into the phase diagram with the regular Fermi liquid and singular-Fermi liquid phases.
\end{abstract}


\end{frontmatter}



\section{Introduction}
\label{intro}

The Kondo effect is a many-body phenomenon in which a localized spin is screened by a cloud of surrounding conducting electrons \cite{Hewson}. The effect manifests itself in electron transport due to an unusual scattering mechanism  and it was observed in metals with magnetic impurities as well as in nanostructures \cite{Kouwenhoven}. In quantum dots the conductance  shows universal dependences \cite{Goldhaber} in agreement with theoretical studies based on a single impurity Anderson model \cite{Pustilnik}.
The problem is obviously richer for multi-dot systems where one can expect interplay of the Kondo ground state with internal magnetic orderings \cite{Georges, Izumida,Bulka2004,Mravlje,Jeong}
as well as a quantum phase transition  \cite{Vojta,Zitko2007,  Roch2008,Hofstetter2007,Logan2,Wang,Galpin}.
There is a competition between the Kondo effect with various intra- and inter-dot electron correlations.
The simplest systems comprising these competitions are two-impurity models which have been comprehensively considered in the literature (see e.g. \cite{Aguado,Sun,Borda,Lopez,Chen,Corgnaglia,Holleitner,Keller2014,Ramsak2} and references therein).

In this paper we are interested in quantum dot trimers for which many aspects have been already studied (see the review \cite{Hsieh2012}). According to Di Vincenzo et al \cite{DiVincenzo2000}  trimers with three electron spins can be good candidates for spin qubits which should be more immune to decoherence processes and may be manipulated by purely electrical pulses. Several groups \cite{laird10,Takakura10,Aers2012,Gaudreau2012} have undertaken experiments to investigate dynamics and coherent manipulations in such systems. An interesting case is a triple quantum dot system with a triangular symmetry (TQD) where spin frustration occurs and the spin entanglement is sensitive to breaking of the triangular symmetry \cite{Ingersent,Kuzmenko,Zitko2008,Oguri2009,Vernek2009,Mitchell2009,Mitchell2010,Luczak12,Urbaniak13}.
Our recent transport studies \cite{Tooski14} concerned a special case of TQD  with three electrons in a TQD and they showed that due to the symmetry breaking the zero-bias conductance changes abruptly from the unitary limit to zero. This effect is driven by a transition between the ground states with different internal spin-spin correlations.

Pustilnik and Glazman \cite{Pustilnik2001} showed that the conductance exhibits a transition from the unitary limit to zero with lowering a temperature which results from interplay of electron scatterings on a system with the triplet and singlet close to degeneracy. This is an evidence of a two stage Kondo effect with the transition from the fully screened Kondo regime at high temperatures to the underscreened $S=1$ Kondo effect at low temperatures. Varma et al \cite{Varma2002} and Mehta et al \cite{Mehta2005} using the renormalization group demonstrated that this is a transition from the regular Fermi liquid (RFL) to the singular Fermi liquid (SFL). At low temperatures, the spin $S$ is partially screened to $S^*=S-1/2$. The residual magnetic moment $S^*$ couples ferromagnetically to the rest of conducting electrons. It was shown also that the scattering matrix tends in a singular manner to the unitary limit \cite{Mehta2005}. For $S=1$ the phase shift is $\delta\approx \pi/2$ plus some singular corrections caused by scatterings on the residual spin $S^*$. A subtle interplay of various scatterings leads to a breakdown of the Fermi-liquid picture.
A fundamental characteristic of the singular Fermi liquid is that
the low-energy properties are dominated by singularities
as a function of energy and temperature.

An interesting experimental exemplification of the underscreened Kondo effect was performed by Parks at al \cite{Parks}. They measured the conductance through individual  cobalt complexes with spin S=1 where controllable stretching of the molecule changed its magnetic anisotropy and induced a transition to the underscreened Kondo regime. Theoretical studies by Cornaglia et al \cite{Corgnaglia} showed that stretching the molecule can also lead to a Kosterlitz-Thouless quantum phase transition from a high-conductance singular Fermi liquid to a low-conductance regular Fermi liquid ground state.

The main purpose of this paper is to consider the problem of electronic
correlations and the role of many-particle states in coherent
transport through the TQD system in all range of electron fillings.
To this end, one first needs to study the isolated TQD, its electronic structures and the ground state features with respect to the local gate potential varying the number of electrons in the system.
Then it is possible to see the condition for
a local moment formation with spin $S=1/2$ and $S=1$ as a prior presumption for the Kondo screening.
Especially we are interested in the quantum phase transition between regular- and singular-Fermi liquid ground states \cite{Zitko2008,Mitchell2009}. The calculations are performed with the numerical renormalization group (NRG), by the NRG-Ljubljana
code \cite{NRG-Ljubljana}.

We also rise the question whether the
quantum phase transition between the different Fermi liquid ground states in the TQD can be understood in
terms of the Friedel-Luttinger sum rule.
For a two-level system, it was already shown \cite{Logan2,Logan2011,Logan2014} that the zero-bias conductance can be expressed only in terms of the dot occupancy according to
a Friedel-Luttinger sum rule, which is applicable to both the screened and underscreened Kondo effects.
Recently \v{Z}itko et al \cite{Zitko2012} predicted an underscreened Kondo effect due to dark states in the parallel double quantum dot system.
We expect similar effects in the triangular TQD system where internal interference processes lead to the Fano resonance and formation of many-body dark states \cite{Emary2007,Poltl2009,Bulka11}.

The paper is organized as follows.
In Sec.\ref{isolated}, electronic structures of the isolated TQD
for all electron fillings along with their corresponding correlators are presented.
The main part of the paper is Sec.3 which presents
numerical results for the TQD coupled with electrodes derived by means of NRG approach for the correlators and conductance. We will show that the system exhibits rich range of regimes with different many-electron states and various local spin orderings which result on the Kondo correlations with conducting electrons.
Finally, in Sec.\ref{conc}, the conclusions are presented with a phase diagram for the regular and singular-Fermi liquid constructed from the analysis of the Luttinger phase changes derived via the Friedel sum rule formula and the conductance.

\section{Isolated triple dots}

\label{isolated}
\begin{figure}[hb]
\centering
\includegraphics[width=0.4\textwidth]{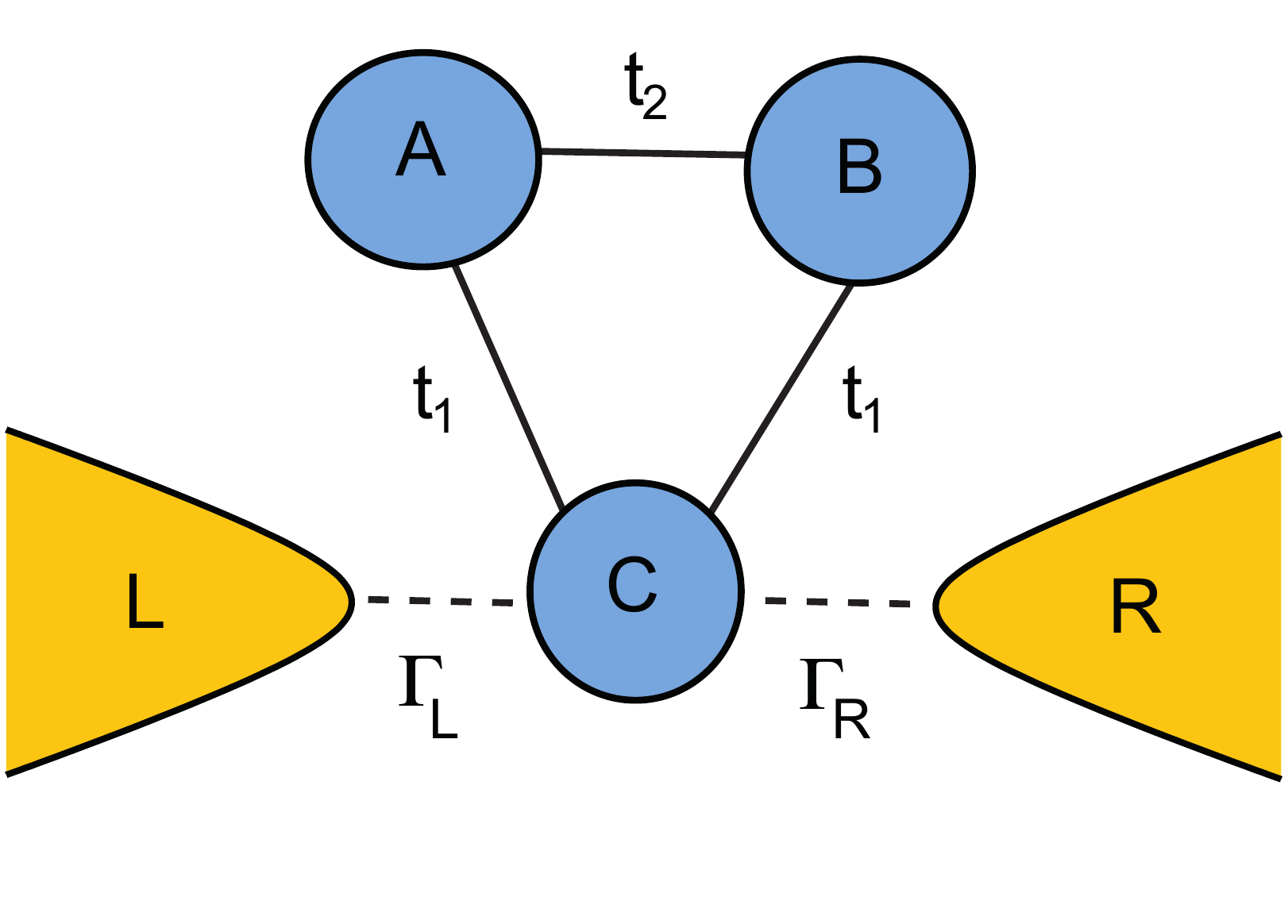}
\caption{Triangular triple quantum dot molecule attached to the leads.}
\label{fig1}
\end{figure}
The considered system of triple quantum dots is presented in  Figure.1. For the isolated TQD the Hamiltonian can be expressed as
\begin{eqnarray}
H_{TQD}=
\sum_{i,\sigma}\epsilon_{i} d_{i\sigma}^\dag d_{i\sigma}+U\sum_{i}  n_{i\uparrow}n_{i\downarrow}+\nonumber\\\sum_{\sigma}[ t_1 d_{C \sigma}^\dag (d_{A \sigma}  +  d_{B \sigma}) +  t_2 d_{A \sigma}^\dag d_{B \sigma} +h.c.]
.
\label{hubbard}
\end{eqnarray}
Here we assume that the size of the dots is small, their intrinsic level spacing is large enough, and therefore one can confine considerations just to a single energy level $\epsilon_{i}=\epsilon$ (for  $i\in\{A,B,C\}$). The second term describes the intradot Coulomb repulsion for two electrons with the opposite spins $\sigma=\uparrow, \downarrow$, where $n_{i\sigma}=d_{i\sigma}^\dag d_{i\sigma}$ denotes  the electron number operator. The last term corresponds to electron hopping between the dots for a symmetric case when the hopping parameters
$t_{CA}=t_{CB}=t_1$ and $t_{AB}=t_2$.

The main purpose of this section is an analysis of electronic correlations in the isolated TQD system for any number of electrons: from zero up to six electrons. We, therefore, derive an electronic structure, find a ground state and all quantities characterizing many body states (as local charges, spin configurations and spin-spin correlations). Numerical results are presented in Fig.2 as a function  of a gate voltage which shifts the position $\epsilon$ of the local levels. We distinguished two cases: weak and strong coupling between the dots A and B (for $t_1>t_2$ and $t_1<t_2$, respectively) for which local charge and spin arrangements are different. Lowering the position of $\epsilon$ we increase the number on electrons in the system, what is seen on the top panels where the charge plots are presented. One can see that the electron-hole symmetry is broken in TQD; there is no mirror symmetry with respect the middle of the figures at $\epsilon+ U/2=0$.
It is worth to mention that in the system one can expect dark states, the states which are decoupled from one of the quantum dot \cite{Michaelis2006,Emary2007,Poltl2009,Bulka11,Luczak12}. For the case $t_1<t_2$ the dark state becomes the ground state for one electron which is equally distributed between the dot A and B, whereas the dot C is empty. Later when we attach electrodes to the dot C, this state becomes decoupled from the electrodes and therefore no current can flow through the system.

When two electrons appear in the TQD they can form a singlet or a triplet state which are mobile (delocalized on three dots). For both considered cases, $t_1>t_2$ and $t_1<t_2$, the triplet has lower energy what is seen in the middle panel for the total spin with $\left\langle \mathbf{S}^2_{tot} \right\rangle=2$. Symmetrically for four electrons the ground state is the singlet with $\left\langle \mathbf{S}^2_{tot} \right\rangle= 0$. In the calculations we take the hopping parameters $t_1$ and $t_2$ positive, but when one changes their sign the position of the triplet and the singlet is exchanged.
In the calculations we take into account a whole space of electron states, including excited states with double electron occupancy, the states which participate in superexchange coupling between the spins. However, for two (four) electrons in TQD  kinetics of charges dominates in the ground state properties and superexchange processes play a minor role \cite{Shim08}.

\begin{figure}[t]
\centering
\includegraphics[width=0.45\textwidth]{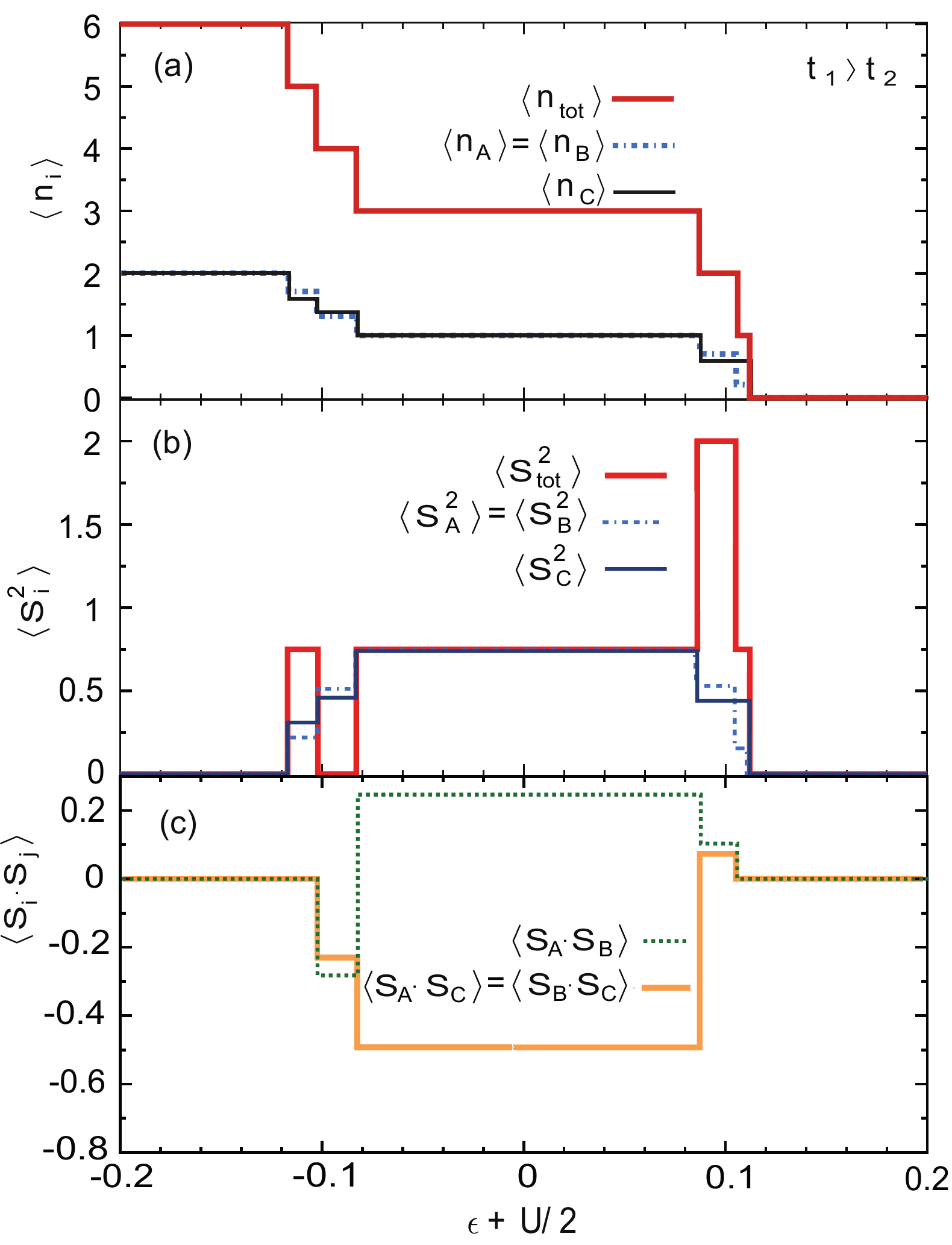}
\caption{\label{fig2}%
Total $\left\langle n_{tot} \right\rangle$ and local $\left\langle n_{i} \right\rangle$  charges (a),
total $\left\langle \mathbf{S}^2_{tot} \right\rangle$ and local $\left\langle \mathbf{S}^2_{i} \right\rangle$ length of spins (b) and
inter-dot spin-spin correlations $\langle \mathbf{S}_i\cdot\mathbf{S}_j\rangle$ (c)
as a function of gate-voltage $\epsilon+U/2$ derived for the isolated TQD with the interdot coupling $t_{1}/D=0.01 > t_{2}/D=0.005$ and strong intradot Coulomb interactions $U/D=0.2$ in units of the half-bandwidth of the conduction band  $D=1$. Notice that in the middle of the plot, for $\left\langle n_{tot} \right\rangle=3$, the ground state is the doublet $|D_T^{S_z}\rangle$ with $\left\langle \mathbf{S}^2_{tot} \right\rangle=3/4$ and ferromagnetic correlations between the spins A and B, $\langle \mathbf{S}_A\cdot\mathbf{S}_B\rangle >0$. The system does not have the electron-hole symmetry: for $\left\langle n_{tot} \right\rangle=2$ the ground state is triplet with $\left\langle \mathbf{S}^2_{tot} \right\rangle=2$ and ferromagnetic correlations between the spins,  $\langle \mathbf{S}_i\cdot\mathbf{S}_j\rangle >0$, while for  $\left\langle n_{tot} \right\rangle=4$ the ground state is singlet with $\left\langle \mathbf{S}^2_{tot} \right\rangle=0$ and antiferromagnetic correlations between the spins,  $\langle \mathbf{S}_i\cdot\mathbf{S}_j\rangle <0$.}
\end{figure}

\begin{figure}[t]
\centering
\includegraphics[width=0.45\textwidth]{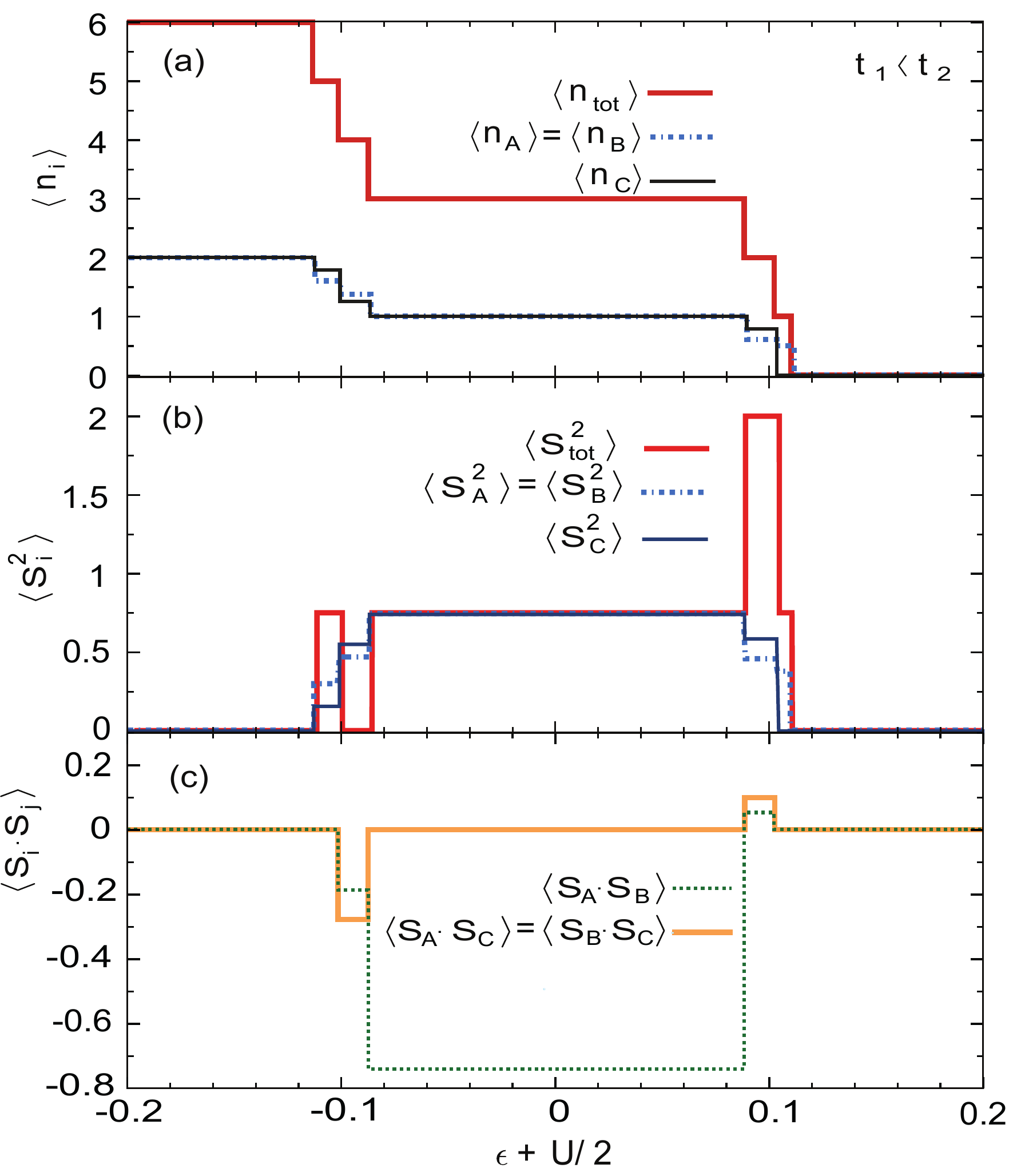}

\caption{\label{fig3}%
The quantities: $\left\langle n_{tot}\right\rangle$,  $\left\langle n_{i} \right\rangle$ (a), $\left\langle \mathbf{S}^2_{tot} \right\rangle$, $\left\langle \mathbf{S}^2_{i} \right\rangle$ (b) and $\langle \mathbf{S}_i\cdot\mathbf{S}_j\rangle$ (c) with respect to $\epsilon+U/2$
for the case $t_{1}/D=0.005 <t_{2}/D=0.01$ for the isolated TQD.
In the middle of the plot, where $\left\langle n_{tot} \right\rangle=3$, the ground state is the doublet $|D_S^{S_z}\rangle$ with perfect entanglement between the spins A and B, $\langle \mathbf{S}_A\cdot\mathbf{S}_B\rangle=-3/4$, whereas the spin C is decoupled, $\langle \mathbf{S}_A\cdot\mathbf{S}_C\rangle=\langle \mathbf{S}_B\cdot\mathbf{S}_C\rangle=0.$ Notice that for $\left\langle n_{tot} \right\rangle=1$ the dark state occurs for which the dot C is empty and the electron is distributed between the dots A and B.}
\end{figure}
In the middle of the plot (around $\epsilon+U/2=0$) the ground state is for three electrons in TQD. Now charge kinetics is suppressed (due to large $U$ and Pauli exclusion principle), therefore superexchange processes dominates \cite{Tooski14,Bulka11,Luczak12,Urbaniak13}.
The ground state is given by doublet states.
Projecting into the singly-occupied subspace, the doublet spin states with $S_z= + 1/2$ can be written as
%
%
%\begin{eqnarray}{lll}
\begin{align}
\label{d1}
|D_S^{1/2}\rangle=&1/\sqrt{2}\;\big(|\uparrow_{A}\downarrow_{B}\rangle-|\downarrow_{A}\uparrow_{B}\rangle \big) \otimes|\uparrow_{C}\rangle,&\\ \label{d2}
|D_T^{1/2}\rangle=&1/\sqrt{6}\; \big[ 2\;|\uparrow_{A}\uparrow_{B}\rangle\otimes|\downarrow_{C}\rangle&
\nonumber\\&- (|\uparrow_{A}\downarrow_{B}\rangle +|\downarrow_{A}\uparrow_{B}\rangle)\otimes|\uparrow_{C}\rangle \big].&
%\end{eqnarray}
\end{align}
It is seen that the state $|D_S\rangle$ is composed from singlet on the AB bond and an electron at the dot C, whereas $|D_T\rangle$ is composed from triplets on the AB bond and an electron with spin $\uparrow$ and $\downarrow$ at the dot C. From Fig.2 one can see that for $t_1>t_2$ the ground state is $|D_T\rangle$ for which the spin-spin correlations at the AB bond are positive,
$\langle\mathbf{S}_A\cdot\mathbf{S}_B\rangle=1/4$,  and $\langle\mathbf{S}_A\cdot\mathbf{S}_C\rangle=\langle\mathbf{S}_B\cdot\mathbf{S}_C\rangle= -1/2$. For the case $t_1<t_2$ one can see maximal entanglement between the spins A and B, $\langle\mathbf{S}_A\cdot\mathbf{S}_B\rangle=-3/4$, whereas the spin C is decoupled,
$\langle\mathbf{S}_A\cdot\mathbf{S}_C\rangle=\langle\mathbf{S}_B\cdot\mathbf{S}_C\rangle=0$.
Here we have an example of monogamy of entanglement \cite{Coffman00}. According to the monogamy concept when two quantum objects, e.g. the spins A and B, are maximally entangled they cannot be
entangled with any third party object.
For a general case when the hopping parameters $t_{AC}\neq t_{BC}$ the ground state is a coherent mixture of the both cases for $|D_S\rangle$ and $|D_T\rangle$.

\section{NRG studies of tripled dot coupled with electrodes}

The numerical renormalization-group is an universal method to study impurity problems, in particular
the Kondo effect, where correct  description of  the screening
of the impurity spin by the conduction-band electrons at low
temperature scales is essential \cite{Krishnamurthy,Krishnamurthy2,NRG-Ljubljana,Hewson,Costi,Bulla2,Hofstetter}.
The essence of the method is a logarithmic discretization of states and a mapping to a one-dimensional
chain Hamiltonian with exponentially decreasing hopping constants which enables
diagonalization iteratively and to keep only the states with the
lowest lying energy eigenvalues. The energy scales are separated since
the matrix elements between the states on vastly different energy scales
are very small and may be neglected.

In this work numerical results were obtained by the NRG calculations
using the discretization and the iteration parameters as in Ref. \cite{Tooski14} where was used
the NRG Ljubljana code
\cite{NRG-Ljubljana},  an implementation of the NRG using
Mathematica and C++. The Mathematica part of the code is used for the
initialization of the problem: using an input of Hamiltonian and
operators of interest in the form of second-quantization expressions,
it automatically generates the eigenvalues and eigenvectors in all
symmetry-adapted subspaces of the full space, as well as the
matrix representations of all required operators. All results presented in this paper
are fully converged, since the three-impurity
single-channel quantum impurity problem studied here
can be analyzed using
relatively modest numerical requirements \cite{Tooski14}.

\subsection{NRG results of correlators}
\label{nrgcorr}

In the following we present and discuss numerical results obtained
by means of the NRG code for the quantities characterising the charge and the spin correlations of the TQD system
as a function of the gate-voltage which shifts the position of  $\epsilon+U/2$.
The TQD system is symmetrically coupled to the leads, $\Gamma_L=\Gamma_R=\Gamma$,
and all results are shown in the strong Coulomb
repulsion regime $U/\Gamma=20$, at a low temperature $T/D=10^{-13}$ where the half-bandwidth of the conduction band  $D=1$ is the largest parameter.

\begin{figure}[t]
\centering
\includegraphics[width=0.45\textwidth]{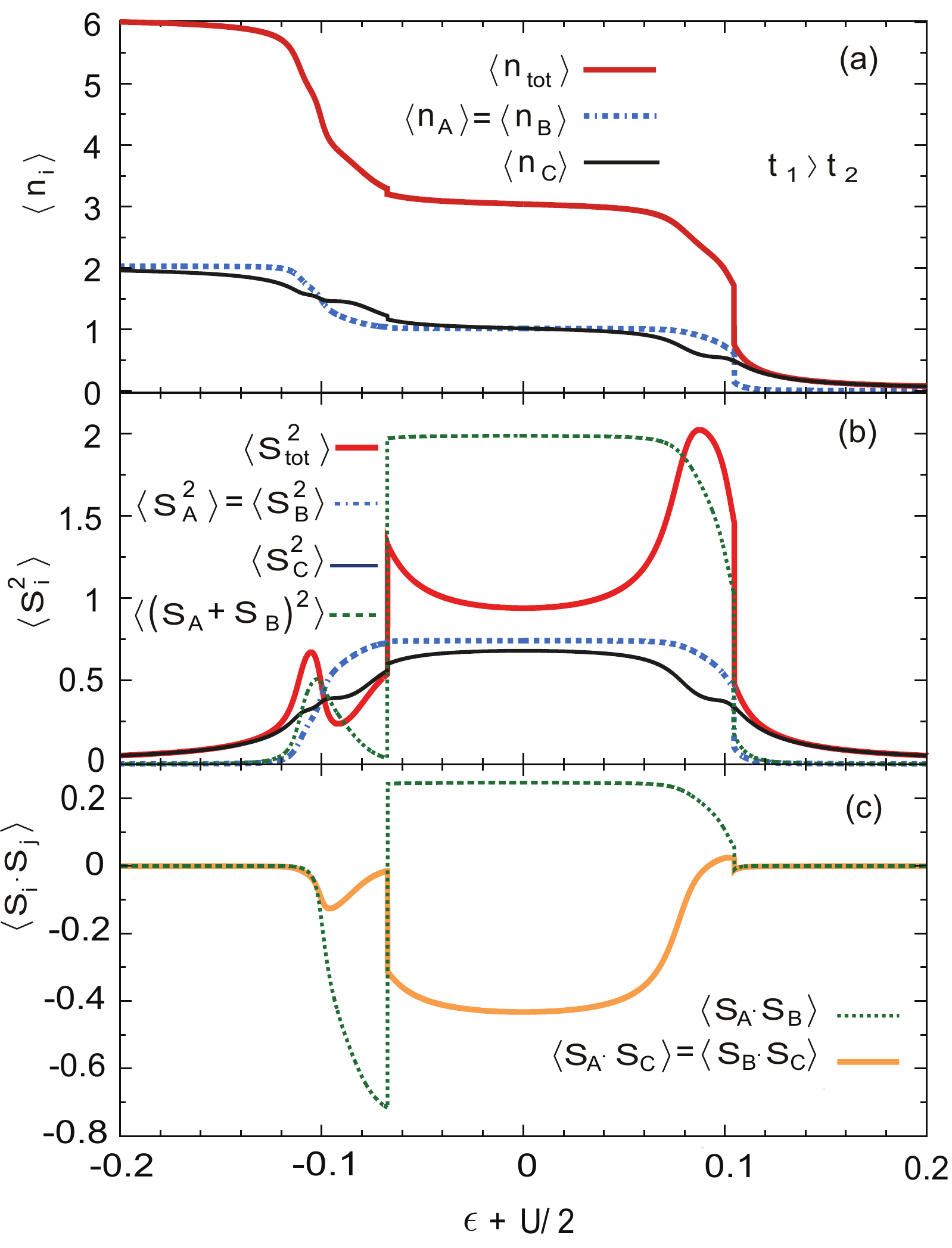}
\caption{The quantities:
$\left\langle n_{tot}\right\rangle$,  $\left\langle n_{i} \right\rangle$ (a), $\left\langle \mathbf{S}^2_{tot} \right\rangle$, $\left\langle \mathbf{S}^2_{i} \right\rangle$ (b) and $\langle \mathbf{S}_i\cdot\mathbf{S}_j\rangle$ (c) derived with respect to  $\epsilon+U/2$ by the NRG method for the case
 $t_{1}/D=0.01 >t_{2}/D=0.005$. The coupling of TQD with the electrodes is taken as $\Gamma/D=0.012$ and $U/D=0.2$. Compared to the low $U/D$ results here transitions between different regimes are gradual except at particular points $\epsilon+U/2 \approx 0.103$ and $-0.07$, the transition from 2 to 3 electron state where the triplet state is formed and the crossover from 3 to 4 electrons, respectively. The formation of local moments is evident from panel (b) and spin-spin arrangements from panel (c).
 }
 \label{fig4}
\end{figure}

\begin{figure}[t]
\centering
\includegraphics[width=0.45\textwidth]{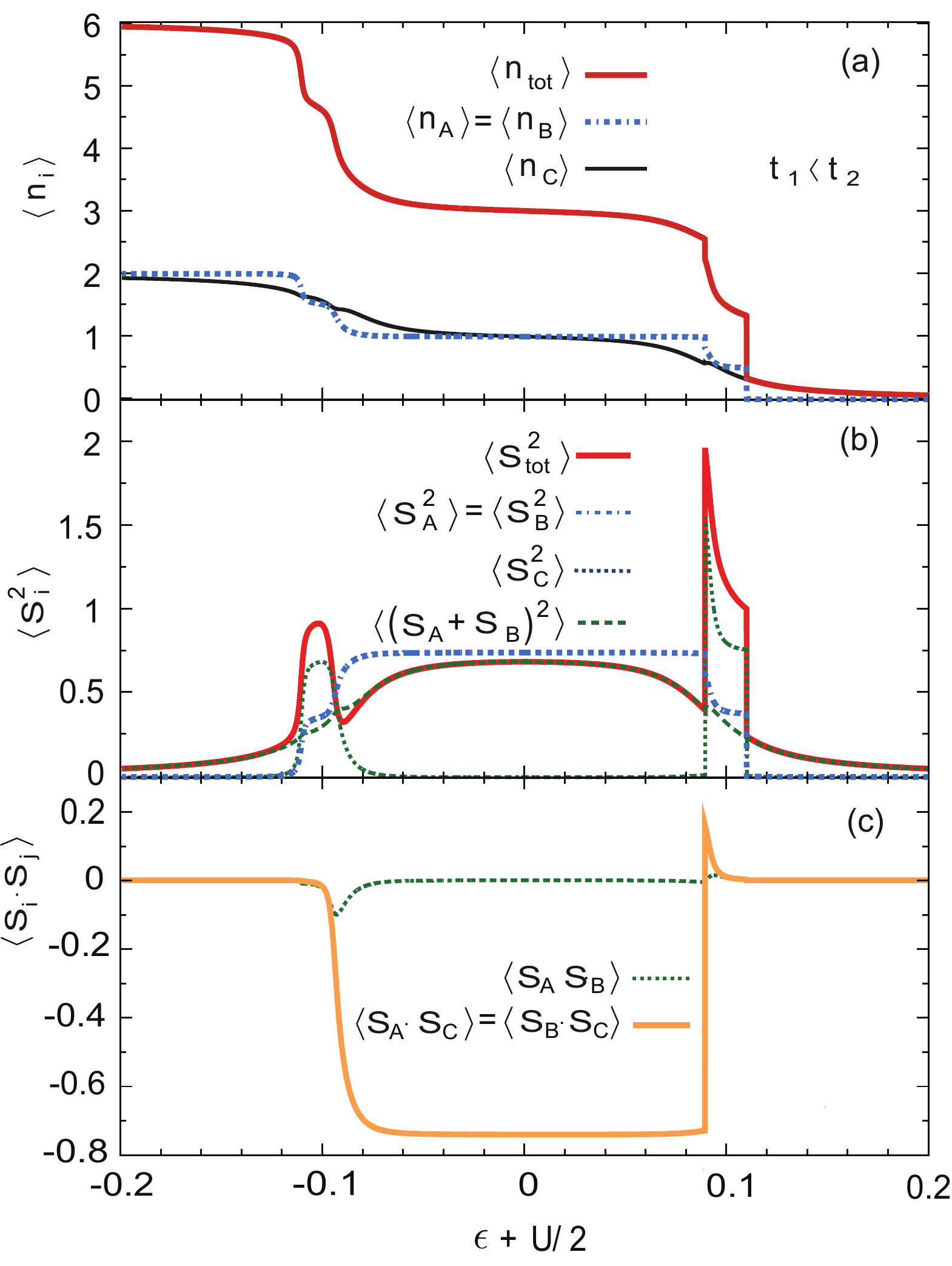}
\caption{The quantities:
$\left\langle n_{tot}\right\rangle$,  $\left\langle n_{i} \right\rangle$ (a), $\left\langle \mathbf{S}^2_{tot} \right\rangle$, $\left\langle \mathbf{S}^2_{i} \right\rangle$ (b) and $\langle \mathbf{S}_i\cdot\mathbf{S}_j\rangle$ (c) derived by the NRG method for the case
 $t_{1}/D=0.005 <t_{2}/D=0.01$. The other parameters are the same as in Fig.\ref{fig4}. Note the sharp transition from 1 to 2 electrons at $\epsilon+U/2 \approx 0.105$ due to the dark state formation, and the transition from 2 to 3 electrons at $\epsilon+U/2 \approx 0.09$. Between these two points the spin of the pair of dots A-B is enhanced, $\left\langle \mathbf{S}_{tot}^2\right\rangle \to 2$.  }
\label{fig5}
\end{figure}

Figs.~\ref{fig4} and \ref{fig5} present the correlators  plotted versus the gate-voltage
for two different cases $t_{1}>t_{2}$ and $t_{1}<t_{2}$, respectively.
Comparing with the isolated TQD, with Fig.\ref{fig2} and \ref{fig3}, now the plots show natural broadening caused by charge fluctuations between TQD and the electrodes.
However one can also see a series of jumps,
which correspond to level crossing between different electron ground states when charging of the TQD system is changed.

\subsubsection{Ferromagnetic case}

Let us first consider the $t_{1}>t_{2}$ case, presented in Fig.~\ref{fig4}, in details.
From the right-hand side the first jumps in the correlators correspond to charging of TQD with two electrons. In Fig.~\ref{fig4}(a) one can see that the plot of
$\langle n_{A} \rangle=\langle n_B\rangle$
shows a sharp jump at $\epsilon+U/2 \approx 0.103$,
when the second electron enters suddenly in the dots $A$ and $B$.  In this situation the triplet state is formed, which is seen in the plots for
$\left\langle \mathbf{S}_{tot}^2\right\rangle$
and $\left\langle \mathbf{S}_i \cdot \mathbf{S}_j\right\rangle$, Fig.~\ref{fig4}(b). The length of the total spin $\left\langle \mathbf{S}_{tot}^2\right\rangle$
achieves its maximal value about 2, and the spin-spin correlations $\left\langle \mathbf{S}_i \cdot \mathbf{S}_j\right\rangle$ are positive. These results suggest an underscreened $S=1$ Kondo effect.

For the case of the half-filled TQD
one can see that $\left\langle \mathbf{S}_{i}^2\right\rangle$
saturates and it is closed to $3/4$, indicating the doublet state with the spin $S=1/2$. Notice that in this voltage range the correlations $\left\langle \mathbf{S}_A \cdot \mathbf{S}_B\right\rangle$ between the spins at the dot $A$ and $B$ are ferromagnetic and the length of the total spin $\langle \left( \mathbf{S}_A+ \mathbf{S}_B \right)^2 \rangle=S_{AB}\left(S_{AB}+1\right)$ reaches its maximal value 2, Fig.~\ref{fig4}b.

The second sharp transitions occur in the plots at
$\epsilon+U/2 \approx -0.07$,
which corresponds to a crossover from the three-electron to the four-electron charge state.
Both of the total spin squares $\langle \mathbf{S}_{tot}^2\rangle$
and $\langle \left( \mathbf{S}_A+ \mathbf{S}_B \right)^2\rangle$
change sharply from their maximal to minimal values,
while the spin-spin correlators $\left\langle \mathbf{S}_A \cdot \mathbf{S}_B\right\rangle$
change abruptly from ferromagnetic to antiferromagnetic.  These results indicate the singlet ground state formation.
In this situation, one can anticipate a quantum phase transition
between the fully screened and the underscreened Kondo effect, which as we will show later can be manifested itself in the electronic transport.

Lowering the gate-voltage more,
$\langle \mathbf{S}_{tot}^2\rangle$ reaches $3/4$ for five electrons in TQD.
This indicates the local moment formation for five electrons in TQD,
where one can expect fully screened $S=1/2$ Kondo effect.

\subsubsection{Antiferromagnetic case}

Let us turn to the case $t_{1}<t_{2}$ presented in Fig.~\ref{fig5}.
The behavior of the average charge
$\left\langle n_{i} \right\rangle$ is similar to the previous case. However, for the case with one electron at TQD one has $\langle n_{A} \rangle=\langle n_B\rangle=0$ due to the dark state formation. Later, at $\epsilon+U/2 \approx 0.105$, when the second electron enters suddenly in the dots $A$ and $B$ all quantities exhibit sharp jumps.
One can also see that $\left\langle \mathbf{S}_{tot}^2\right\rangle$
reaches its maximal value about $2$ when two electrons form the mobile triplet state.
In this situation one can expect the underscreened $S=1$ Kondo effect
for both the ferromagnetic and antiferromagnetic cases.
The region with the triplets is separated by two sharp changes in the plots on the right-hand side of the figure which
corresponds to fluctuations between the one-electron and
the two-electron states at $\epsilon+U/2\approx 0.105$,
as well as between the two-electron and the three-electron
states at $\epsilon+U/2 \approx 0.09$.
We show later that the second sharp transition is related to the quantum phase transition between the underscreened and fully screened Kondo effect.
In the middle of the plot there is a region with three electrons in TQD, where  $\left\langle \mathbf{S}^2_{i}\right\rangle$
is closed to $3/4$. In this region, $\langle \left( \mathbf{S}_A+ \mathbf{S}_B \right)^2 \rangle$
is near zero and $\left\langle \mathbf{S}_A \cdot \mathbf{S}_B\right\rangle\approx -3/4$ due to the formation of the singlet state between dots $A$ and $B$. The spin at the dot $C$ is unentangled with the others and it can form a Kondo cloud with spins of conducting electrons.
The behavior of four and five electrons is similar to the ferromagnetic case.

\subsection{NRG studies of transport}
\label{nrgtrans}

Rich structure of possible states of the TQD system discussed in Section A reflects in the conductance properties. One of the aims of this paper is also a quantitative analysis of the Friedel-Luttinger sum rule (FLSR), which connects the conductance by  the occupancy of the system and the Luttinger integral, as discussed in detail in Refs. \cite{Logan2,Logan2011,Logan2014}. In particular, the zero-bias conductance at zero temperature is given by
\begin{equation}
\label{eq:FSRG}
G={G_0} \sin^2\delta(\epsilon_F),
\end{equation}
where $G_{0}=2e^2/h$ is the conductance quantum, the phase shift $\delta(\epsilon_F)$ depends on the occupation of  the impurity $n_{imp}$ and the Luttinger integral $I_L$,
\begin{eqnarray}
\label{eq:FSR39}
\delta(\epsilon_F)=\frac{\pi}{2}n_{imp}+I_L.
\end{eqnarray}
Luttinger and Ward  \cite{Luttinger2} argued that $I_L=0$ for Fermi liquid systems, in our case therefore
$I_L = 0$ in the regular Fermi liquid (RFL) phase,
whereas $\left| I_L \right|= \pi/2$ in the singular Fermi liquid (SFL) phase.
As such, the Luttinger integral can be the hallmark of both
the RFL and the SFL phases in a rather deep sense.
These results should have an immediate
consequence on the zero-bias conductance $G$ in both of the phases.
Therefore, we analyze the TQD system in details by
comparing the conductance calculated directly form the spectral function
with the conductance deduced from the Friedel-Luttinger sum rule.
This can be considered as a way to determine the corresponding phase
for each of the ground states. As discussed later, the detail investigation of the conductance presents the ground state characteristics
and their corresponding quantum phase transitions
which separate the RFL and the SFL ground states.

\begin{figure}[t]
\centering
\includegraphics[width=0.47\textwidth]{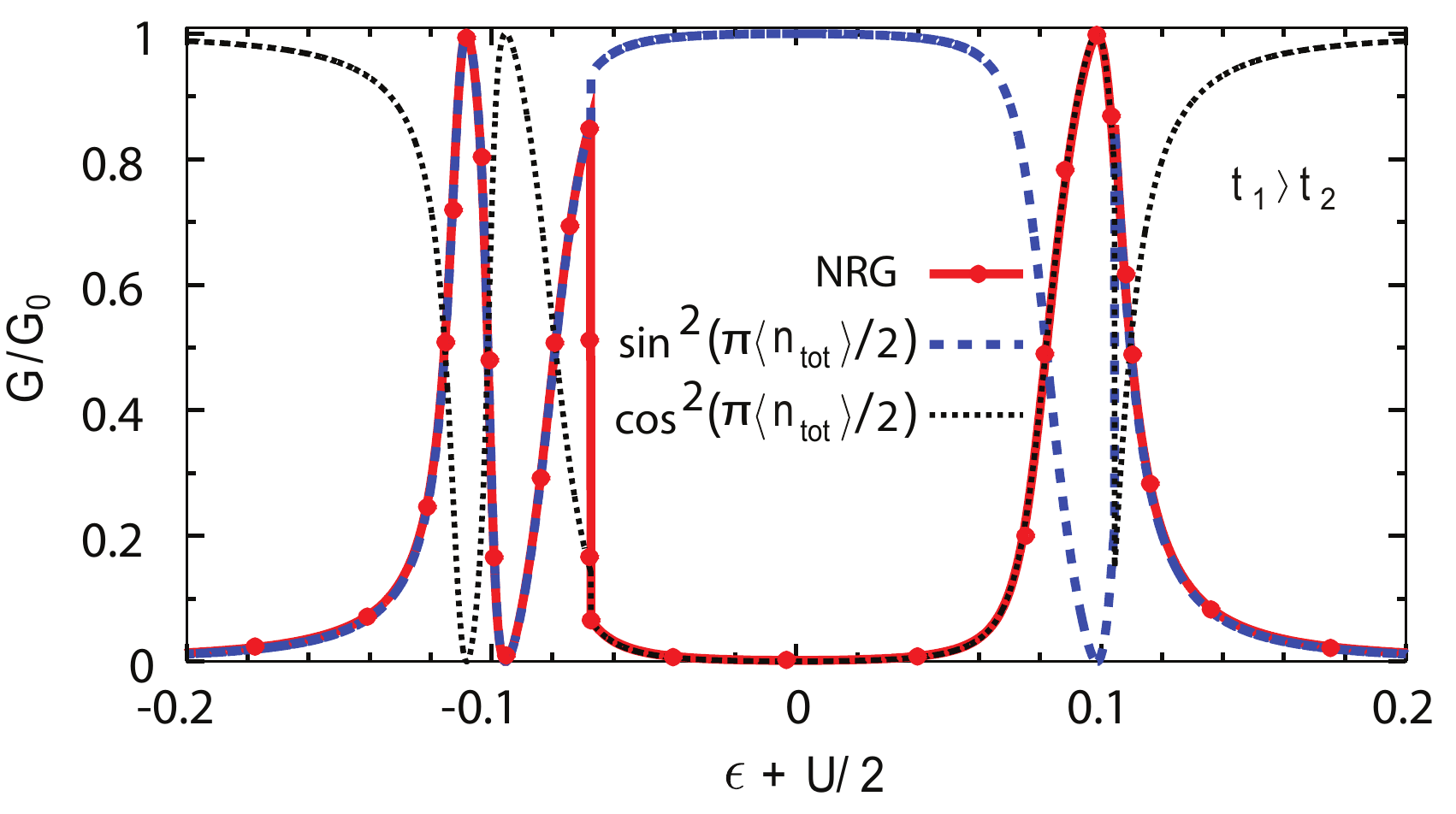}
\caption{The conductance $G$ calculated directly form the spectral function by NRG method
as a function of gate-voltage $\epsilon+U/2$ for $U/D=0.2$,
$\Gamma/D=0.012$, $t_{1}/D=0.01$, and $t_{2}/D=0.005$.
It is compared with the conductance deduced
from the Friedel-Luttinger sum rule, Eq.(\ref{eq:FSRG}),
to indicate the phases of the system.}
\label{fig6}
\end{figure}

\subsubsection{Ferromagnetic case}

With the reference to the analysis of correlation functions we present results for two distinct regimes. For the $t_{1}>t_{2}$ case
Fig.~\ref{fig6} presents the gate-voltage dependence
of the conductance calculated by the NRG. Here at $\epsilon+U/2 \approx 0.11$
the system undergoes a transition from one-electron ground state with the total spin $S=1/2$
to two-electron ground state with the total spin $S=1$
where the underscreened Kondo effect is expected, as known from the results for the corresponding correlators in Fig.~\ref{fig4}.
At this point, a Fano resonance reflects with a small sharp peak and deep in $G$.
Interesting problems arise due to the screening of the magnetic moment
$S=1$ for the two-electron ground state of the TQD by conduction band electrons.
The question is, whether the Kondo effect appear for the TQD system with two electrons,
and if so, how is it related to the ground state of the isolated system? We apply the FLSR formula and compare the results with
$G$ obtained directly from the NRG. To be specific, we observe that the conductance satisfies the relation
$G/G_{0}=\sin^2(\pi \left\langle n_{tot}\right\rangle/2)$ for one-electron ground state or $G/G_{0}=\cos^2(\pi \left\langle n_{tot}\right\rangle/2)$ for the two-electron ground state case. This means that TQD as a whole behaves as a magnetic impurity.

We take here $\left\langle n_{tot}\right\rangle$ as the total number of electrons in TQD
which is equal to $n_{imp}$ in Eq. \ref{eq:FSR39}. The results perfectly agree with one or the other case which
confirms also that the FLSR is satisfied for the total number of electrons rather than the local number of electrons, {\it i.e.} the number of electrons at the dot $C$.

In section 2, it has been shown that one- and two-electron ground states
are qualitatively different, each corresponding to  a different spin configuration.
Therefore, these states are expected to be separated
by a quantum phase transition.
Moreover, the conductance reaches its maximal value $G/G_{0}=1$ when
there is two electrons in TQD.
In fact, the screened and underscreened Kondo regimes can often be
differentiated via their conductance \cite{Roch2008}.

We note also that the conductance smoothly
crosses from two-electron to three-electron ground state, without a quantum phase transition.
This is already expected from the smooth transition for the corresponding correlators,
presented in Fig.~\ref{fig4}.
In the previous chapter it has shown that for the $t_{1}>t_{2}$ case,
three-electron ground state is a doublet
with the total spin $S = 1/2$.
In this situation, there should be some kind of the Kondo effect
due to the presence of local magnetic moment $S = 1/2$.
Since the total spin is $S = 1/2$, one should obtain the usual single-impurity Kondo effect
where $G/G_{0}=\sin^2(\pi \left\langle n_{tot}\right\rangle/2)$ is expected.
However, this situation does not occur and the conductance satisfies
the relation $G/G_{0}=\cos^2(\pi \left\langle n_{tot}\right\rangle/2)$,
indicating the underscreened Kondo effect.

\begin{figure}[t]
\centering
\includegraphics[width=0.47\textwidth]{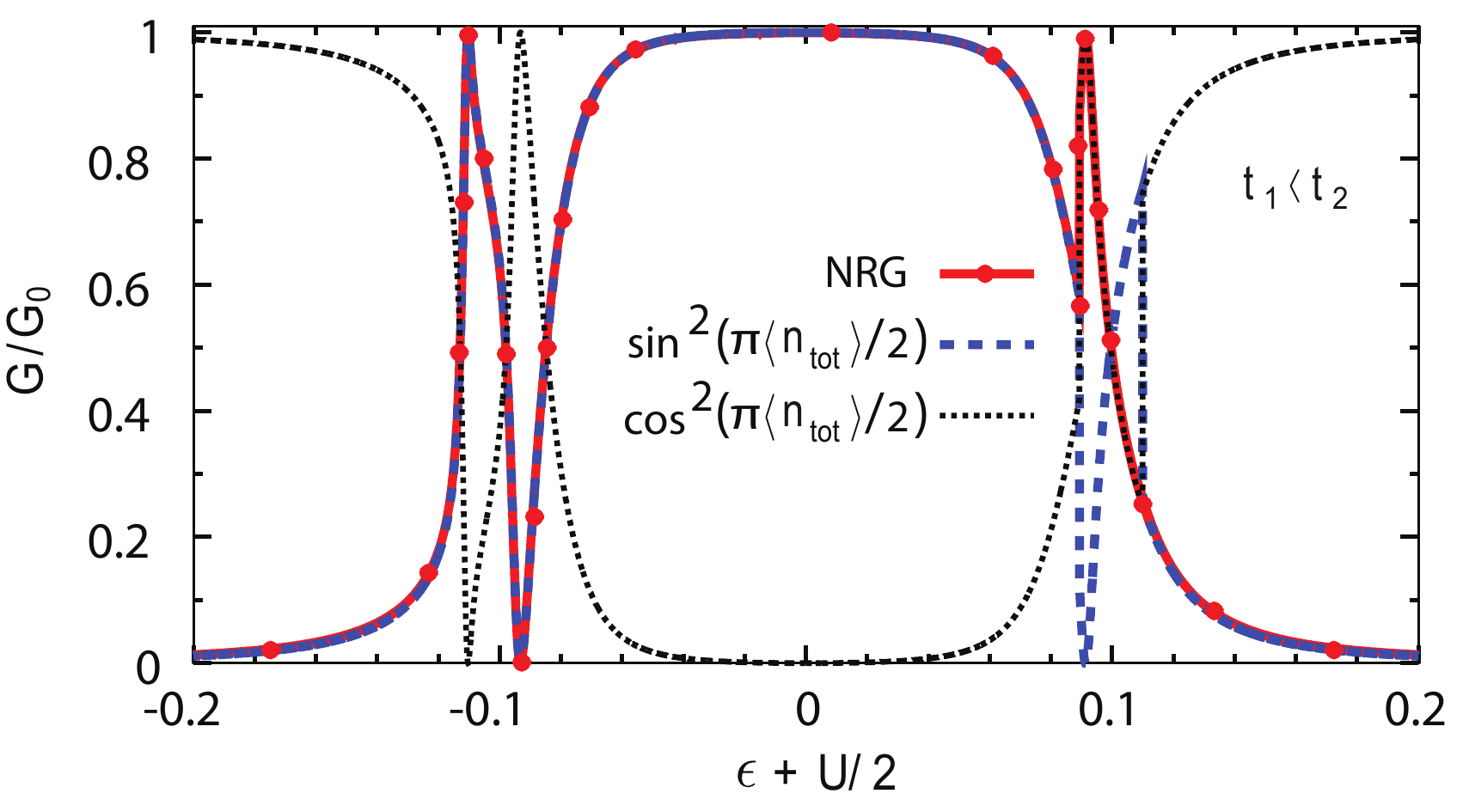}
\caption{The conductance $G$ calculated directly form the spectral function by NRG method
as a function of gate-voltage $\epsilon+U/2$ for $U/D=0.2$,
$\Gamma/D=0.012$, $t_{1}/D=0.005$, and $t_{2}/D=0.01$.
It is compared with the conductance deduced
from the Friedel-Luttinger sum rule, Eq.(\ref{eq:FSRG}),
to indicate the phases of the system.}
\label{fig7}
\end{figure}
Next we observe a sharp transition in $G$ from $0$ to $0.8\;G_{0}$
at $\epsilon+U/2 \approx -0.07$. It is related to the level-crossing between
three- and four-electron ground states (see the correlators in Figs.~\ref{fig2} and \ref{fig4}).
The transition in $G$
finds its counterpart in the jumps of the total impurity occupancy
$\left\langle n_{tot}\right\rangle$ and the spin square $\left\langle \mathbf{S}^2_{tot}\right\rangle$
as well as the spin-spin correlators $\left\langle \mathbf{S}_i\cdot\mathbf{S}_j\right\rangle$.
It is remarkable that the transition
occurs precisely at the point where
$\langle \left( \mathbf{S}_A+ \mathbf{S}_B \right)^2 \rangle$
changes from $2$ to $0$,
signaling the transition in the local moment  from $S_{AB}=1$ to $0$, see Fig.~\ref{fig4}.
This is again the transition between the SFL and the RFL phase exhibited
with a sharp transition from $G/G_{0}=\cos^2(\pi \left\langle n_{tot}\right\rangle/2)$ to
$\sin^2(\pi \left\langle n_{tot}\right\rangle/2)$
due to the fundamentally distinct ground states. This effect should be detected in experimental observation of the conductance.

As already shown in section 2 the
four-electron ground state is a singlet
with no local moment,
whereas five-electron ground state
is a state with $S=1/2$.
This should lead to the valley in the conductance for four-electron ground
state where $G\approx 0$  - as it has been already expected.
However, $G$ reaches its maximal value
when there is precisely five electrons in the TQD.
This is manifestation of the typical Kondo effect with the fully screened spin $S=1/2$.
With a further lowering of the gate-voltage the system enters into the mixed valence regime and
and finally to the full-orbital regime (with six electrons) where $G$ is reduced to zero.

Considering the case with $t_{1}>t_{2}$ we summarize that
the ground state of TQD is the SFL for two and three electrons,
whereas for the other electron fillings the ground state is the RFL.

\subsubsection{Antiferromagnetic case}

As it is known from the results for correlation functions, Fig.~\ref{fig5}., for the $t_{1}<t_{2}$ case the structure of states changes. In Fig.~\ref{fig7} is presented conductance in this regime, as a function of the gate-voltage, with the same values of parameters as in Fig.~\ref{fig6}.
For one electron,
the system can be in the dark state where the probability to
find the electron in dot $C$ is zero, and consequently the transport through the
dot \cite{tooski3} is blocked, in agreement with the earlier results \cite{Michaelis2006,Emary2007}.
As in the previous case appears the quantum phase transition
between the RFL phase to the SFL phase
at the crossing point $\epsilon+U/2\approx 0.11$ from one-electron to two-electron ground state.
In this case the transition is signaled by a change of the conductance from the relation
$G/G_{0}=\sin^2(\pi \left\langle n_{tot}\right\rangle/2)$
to $G/G_{0}=\cos^2(\pi \left\langle n_{tot}\right\rangle/2)$.
At $\epsilon+U/2 \approx 0.092$, another quantum phase transition
takes place, from the SFL to the RFL ground state, which manifests in the conductance change
from $G/G_{0}=\cos^2(\pi \left\langle n_{tot}\right\rangle/2)$
to $G/G_{0}=\sin^2(\pi \left\langle n_{tot}\right\rangle/2)$.
This transition occurs at the level-crossing between
the two- and three-electron ground states.
A sharp transition in $G$ is accompanied by jumps in its counterparts, namely
in the $AB$ correlators $\left\langle \mathbf{S}_A\cdot\mathbf{S}_B \right\rangle$, from a ferromagnetic coupling with $\left\langle \mathbf{S}_A\cdot\mathbf{S}_B \right\rangle=0.2$ to an antiferromagnetic coupling $\left\langle \mathbf{S}_A\cdot\mathbf{S}_B \right\rangle=-0.75$,
as well as in $\left\langle \mathbf{S}^2_{tot}\right\rangle$ (from $2$ to $0.4$).

In the previous chapter has been shown that
in three-electron case the spins on the dots $A$ and $B$ form a singlet
with $\langle \left( \mathbf{S}_A + \mathbf{S}_B \right)^2 \rangle =0 $.
In this situation, one gets the same conductance as in
the usual $S=1/2$ single impurity Kondo effect
due to the decoupling of the dots $A$ and $B$ from the central dot
$\left\langle \mathbf{S}_A\cdot\mathbf{S}_C \right\rangle
=\left\langle \mathbf{S}_B\cdot\mathbf{S}_C\right\rangle=0$,
see Fig.~\ref{fig5}.
At $\epsilon+U/2 \approx 0.07 $
the system undergoes a crossover from three- to four-electron ground state
without a quantum phase transition what is in contrast to the $t_{1}>t_{2}$ case.
This has been already expected from the smooth evolution of the correlators (see Fig.~\ref{fig5}).

The behavior of the conductance in the region of four- and five-electron ground states
is similar to the $t_{1}>t_{2}$ case considered in the previous section.

\section{Summary}
\label{conc}

The TQD molecule considered in this paper is an interesting example since in a single system only by changing the chemical potential or local bias one can sweep through a rich range of different spin configurations regarding the whole system or only the subsystems. Two distinct cases are considered here, characterized as (anti)ferromagnetic with respect to the subsystem A-B, each exhibit specific spin entanglement. The evolution of
the conductance can be explained by the Friedel-Luttinger sum rule
which is applicable to both the regular- and singular-Fermi liquid phases.
The FLSR relates the conductance to the impurity charge and the Luttinger integral.
It has been confirmed numerically that the Luttinger integral takes a
value characteristic to the quantum phases of the system, i.e. $I_L=0$
in the regular-Fermi liquid (RFL) phase
and $I_L=\pi/2$ in the singular-Fermi liquid (SFL) phase.

\begin{figure}[t]
\centering
\includegraphics[width=0.45\textwidth]{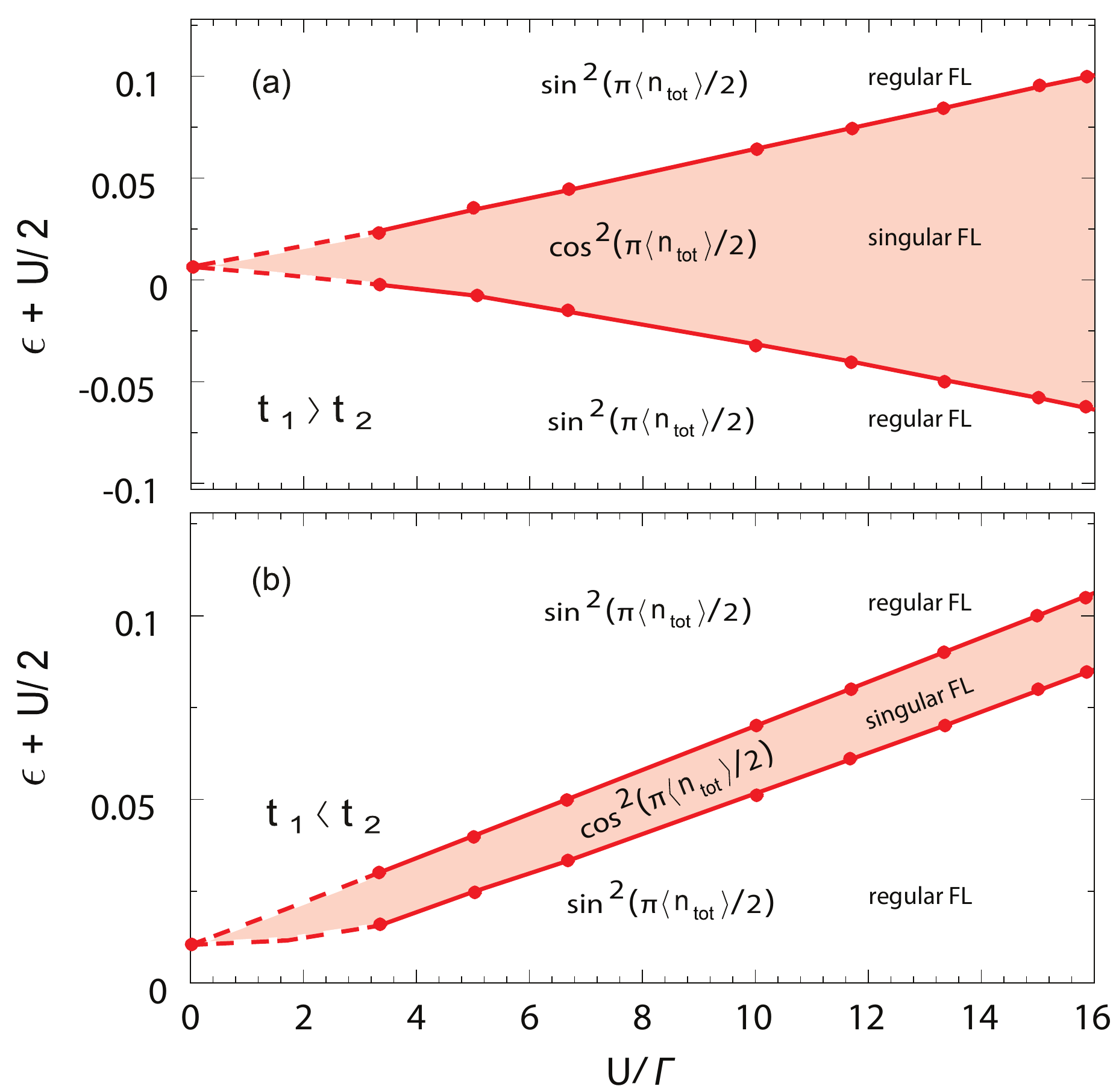}
\caption{
Phase diagram obtained by means of the NRG code for the TQD molecule coupled with the electrodes, $\Gamma/D=0.012$, where solid curves separate regular Fermi liquid
and singular Fermi liquid regions. Fig.a presents the case with $t_{1}/D=0.01$ ,  $t_{2}/D=0.005$, while Fig.b the case: $t_{1}/D=0.005$ and $t_{2}/D=0.01$. }
\label{fig8}
\end{figure}

\begin{figure}[t]
\centering
\includegraphics[width=0.45\textwidth]{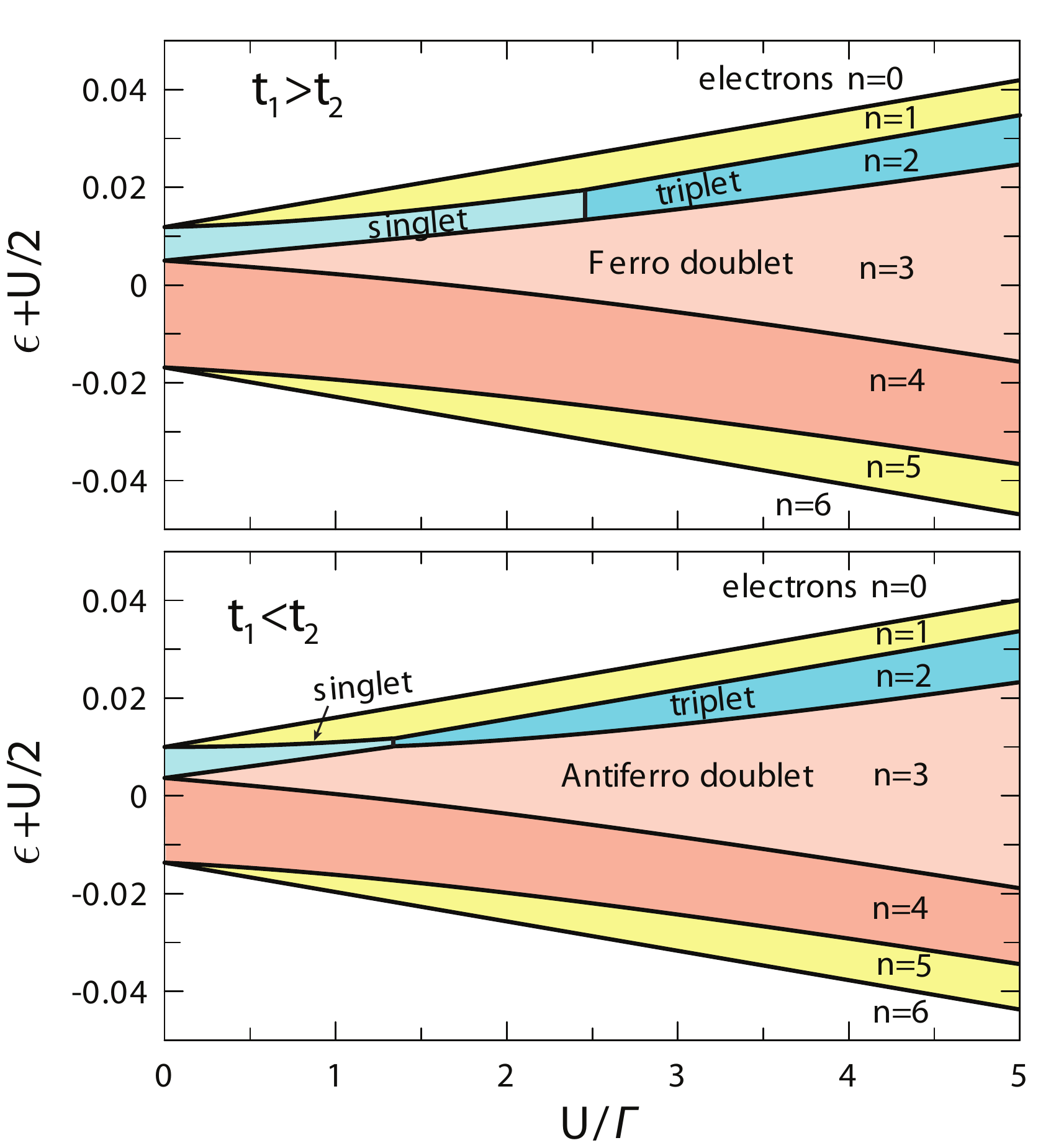}
\caption{\label{diagr-iso}%
Phase diagram
for the isolated TQD molecule showing various charged states for (a): $t_{1}/D=0.01$ and $t_{2}/D=0.005$, (b): $t_{1}/D=0.005$ and $t_{2}/D=0.01$.}
\label{fig9}
\end{figure}

Our results are summarized in Fig.\ref{fig8} which presents the phase diagram for the RFL and SFL. As a function of $U/\Gamma$ is shown the bias regime $\epsilon+U/2$ where the sine- or cosine- conductance relation indicates RFL of SFL regimes, respectively. Bullets represent precise values of the crossover as determined by careful NRG analysis and lines represent the guide to the eye only.
Note that the results for conductance presented in Fig.\ref{fig6} and \ref{fig7} correspond to strong Coulomb interactions with $U/\Gamma =16$. The phase diagrams in Fig.\ref{fig8} should be considered together with Fig.\ref{fig9} showing internal orderings in the isolated TQD for all electron fillings. The SFL phase has been detected when the triplet state with $S=1$ is formed in the TQD molecule. SFL appears also for the case $t_1>t_2$ with three electrons when the ground state is $|D_T^{1/2}\rangle$,  Eq.(\ref{d2}). In this situation there is a specific ferromagnetic coupling between spins in the A-B subsystem which manifests itself in conductance as well.

It is worth to notice that in the TQD system the ground state appears with the total spin $S=0$, 1/2 and 1, which can be understood by the Hund's rules \cite{Korkusinski2007}. For heavy atoms higher orbital states can be degenerate but Coulomb and spin-orbit interactions remove the degeneracy, and due to the Pauli principle a high spin ground state becomes favorable.   In the TQD with the perfect triangular symmetry and in the absence of electron interactions two orbital states with opposite wave vectors are degenerate. Fig.\ref{fig9} shows the ground state diagram for the isolated TQD with respect to the intra-dot Coulomb interactions $U$ and for all electron fillings. For two electrons and small $U$ the ground state is  the singlet, but it can be changed to the triplet for $U/\Gamma \lesssim 2.4$ and $U/\Gamma \lesssim 1.4$ for $t_1>t_2$ and $t_1<t_2$, respectively. Unfortunately we could not detect such the transition in the NRG calculations in this regime. We observed a sharp transition of conductance with $\epsilon$, when the electron number $n$ changes $1 \rightarrow 2 \rightarrow 3$, but the transition range was too small to distinguish difference in the sine- or cosine-relations.

 \section*{Acknowledgements}
We acknowledge the support by  National Science Centre (Poland) under the contract DEC-2012/05/B/ST3/03208 (S.B.T. and B. R. B.) and
Slovenian Research Agency Grant No. P1-0044 (A.R.)



\section*{References}
\bibliographystyle{elsarticle-num}

\bibliography{references}

%\begin{thebibliography}{00}


%\bibitem{}

%\end{thebibliography}

\end{document}